\documentclass[12pt,a4paper,twoside,final,notitlepage, leqno]{article}
\usepackage[english]{babel}
\usepackage[T1]{fontenc}  
\usepackage{epsfig, graphicx, subfigure, amssymb}
\usepackage{epsfig, graphicx, amssymb}
\usepackage{xfrac}
\usepackage{amsmath,amsthm,epsfig,amsfonts,bbm}  
\usepackage{float}
\usepackage{color}
\def\br {\break}

\newcommand{\monitem}{ \smallskip \noindent $\bullet$ \quad  } 
\newcommand{\moneq}{\vspace*{-7pt} \begin{equation} \displaystyle } 
\newcommand{\moneqstar}{\vspace*{-6pt} \begin{equation*} \displaystyle } 
\newcommand{\monendstar}{\vspace*{-6pt} \end{equation*}   }
\newcommand{\monend}{\vspace*{-7pt} \end{equation}   }

\def\section*#1{}

\usepackage{fancyhdr}
\renewcommand{\headrulewidth}{0pt}

\begin{document} 

\fancypagestyle{plain}{ \fancyfoot{} \renewcommand{\footrulewidth}{0pt}}
\fancypagestyle{plain}{ \fancyhead{} \renewcommand{\headrulewidth}{0pt}}

~


\centerline {\bf \LARGE  On Macroscopic Intricate States  }

 \bigskip  \bigskip \bigskip

\centerline { \large    Fran\c{c}ois Dubois$^{ab}$ }

\smallskip  \bigskip

\centerline { \it  \small   
$^a$  Conservatoire National des Arts et M\'etiers, LMSSC,  Paris, France.} 

\centerline { \it  \small  $^b$   Afscet (French Association for Systems Science) }


\bigskip  

\centerline {02 decembre   2017 
{\footnote {\rm  \small $\,$ 
Contribution presented at the  ``WOSC 2017''  Conference, Roma (Italy),  25 - 27 January 2017. 
Published in {\it Kybernetes}, volume 47, pages 321-332, 
december 2017.}}}


\bigskip  
\noindent {\bf \large Abstract}

The present contribution is in the field of 
quantum modelling of macroscopic phenomena. The focus is on one  enigmatic
aspect of quantum physics, namely the Einstein-Podolsky-Rosen paradox and entanglement. 
After a review of the state of the art concerning macroscopic quantum effects and
quantum interaction, this contribution proposes a link between embryology and acupuncture 
in the framework of macroscopic intricate states induced by quantum mechanics. 
%
%
Could a weak form of intrication be maintained  during stem cell division in order 
to interpret the acupuncture meridians as an explicit manifestation of a macroscopic intricate system?  
%
%
%
%



\bigskip \bigskip   \noindent {\bf \large    Introduction }   

\smallskip
First recall  the astonishing quantum phenomenon observed by Aspect and his colleagues 
(Aspect, Grangier, Roger, 1982):  
using calcium cascade source,  a double photon is emitted in two opposite directions. 
From the quantum point of view, this double photon is a unique pure state, 
an entangled state, even if
it is located in two different places.
When a measure is done on one of the photons, the reduction of the wave packet operates instantaneously
on both components. 
 From a so-called classical realistic point of view, 
we are in presence of two objects interacting together whereas from the quantum point of view, 
we are in front of a unique  entangled state 
 that occupies two different macroscopic spatial positions 
during the Einstein-Podolsky-Rosen-Bohm experiment (Einstein, Podolsky and Rosen, 1935, 
Bohm, 1951).  When a measure occurs, 
the so-called reduction of the wave packet of the quantum approach predicts that the 
entangled state remains unique and responds in a holistic manner even if it occupies 
two separate space positions! There is a natural problem with the confrontation 
of such a point of view with the Einsteinian realism and in particular the theory of 
relativity that claims that no interaction can proceed at a celerity superior to the 
one of the light.

\smallskip \noindent  
A detailed analysis of possible cross-correlations has been proposed 
by Bell (1964). 
As a result, the so-called Bell inequalities show that precise experiment is possible 
in order to test whereas the two components of the entengled 
state remains correlated 
or not when they occupy different space positions. The experiment of two entangled 
photons has been proposed and realized with a great success by Aspect. 
The result shows that quantum mechanics gives the good prediction; the Bell 
inequalities are not satisfied by the experiment, even if 
``in many other situations, the Bell inequalities are not violated'' (Aspect, 2005). 
Consequently, the holistic vision of the entangled photons is now experimentally 
well established. 

\smallskip \noindent 
Nevertheless, a main difficulty of these micro-physics experiments is due 
to the so-called decoherence, modelized by Zurek (1982) and experimentally 
established by Haroche and his co-workers (Brune {\it et al.}, 1996). 
When interacting with the environment, mesoscopic quantum systems loose 
quickly their coherence properties. 

\smallskip \noindent 
In this contribution, we are interested in  possible macroscopic entangled states.
In a first section, we review the main facts concerning this subject, 
including quantum  key distribution    and the quantum computer. 
We develop in a second part the fractaquatum hypothesis, 
motivated by the remark that  Nature is both fractal and quantum. 
The fractaquantum hypothesis express that the quantum approach is relevant 
for all the elements  in Nature, whatever their size. 
With this fractaquantum framework, we exlore the possibility of intricate states in 
biology. 
In particular a link between acupuncture and embryology in Section~3. 

\fancyhead[EC]{\sc{ Fran\c cois  Dubois }} 
\fancyhead[OC]{\sc{On Macroscopic Intricate States}} 
\fancyfoot[C]{\oldstylenums{\thepage}}

\bigskip \bigskip   \noindent {\bf \large    1) \quad  Macroscopic entangled states }   

First observe a distinction between ``entangled states'' and ``intricate states'' 
as in our title. 
The term ``intrication''  seems clear, referring to a
process that leads to entanglement. Examples in the literature wants to reserve the term
``intricate states'' to ``localised entanglement'', 
such as the intrication happening when a
microscopic particle interacts with a measuring device, for instance when an alpha-particle
enters a detector.  
In a strictly point of view, the photon pairs in the Aspect's experiment constitute a 
macroscopic entangled state. But such phenomena are also associated to quantum computers, 
quantum key distribution and  quantum interaction.

\monitem   Quantum computer    

\noindent 
The possibility of development of 
a  quantum computer has been suggested by Feynman (1982). 
A $ \, \pm \,$ bit is replaced by a qubit, mathematically a set of two complex numbers
whose sum of square modulus is equal to  one. 
From a physical point of view, 
a quantum computer is a true macroscopic structure at a mesoscopic level 
and we refer the reader to the book of Nielsen and Chuang (2000). 
A quantum computer uses physics to perform an infinite number of scalar operations
per  cycle. 
The Shor algorithm (Shor, 1994) establishes that a quantum computer 
can factorize any integer  $N$ with a very fast algorithm, 
with O($ (\log N)^3$) elementary operations.
A first factorization ($15=3\times 5 $) was obtained quickly (Vandersypen {\it et al.}, 2001).
The progress are  thereafter relatively slow~:  
factorization ($143=11\times 13 $) with a 4-qubits computer 
(Xu {\it et al.}, 2012).
The  first reprogrammable quantum computer is constructed by Debnath {\it et al.} (2016). 
Observe that the decoherence is a strong limitation to the development of quantum computer. 
The  work of Ofek {\it et al.}  (2016) show the state of the art on the
of quantum error correction.

\monitem   Quantum key distribution    

\noindent 
The communication of private  keys is a fundamental question in cryptography.
An important breakthrough has been obtained in 1984 with the ``BB84'' protocol, 
by taking into account the non-commutation of operators with  oblic polarizers 
(Bennett and Brassard, 1984). 
The research  group of Gisin has applied this protocol  under Lake Geneva  (Muller  {\it et al}, 1996) 
and extended it some years later  (Branciard {\it et al.}, 2005).
It has been also implemented over 80~km of optical fibre by the group of Grangier 
(Jouguet  {\it et al.}, 2013). 
The violation of Bell's inequalities is also a security test of a cryptographic installation 
against a possible agression. 
The Ekert protocol (Ekert, 1991) uses explicitly the violation of Bell's inequalities
with an  entangled pairs of photons. It can be viewed as an industrialization of the Aspect experiment. 
In a recent contribution (Vazirani and Vidick, 2014), 
a proof of security of a slight variant of Ekert's original entanglement-based protocol
is presented.  

\monitem   Quantum interaction       

\noindent 
We give in this sub-section some decorrelated discoveries showing the 
important scientific activity in the field. 
A Sino-German team has  teleported quantum information from one ensemble of atoms 
to another 150 metres away (Bao {\it et al.}, 2012). 
The classification of macroscopic quantum effects is difficult. 
Farrow and Vedral (2015) propose three not mutually exclusive classes defined by mass, 
with interference of macromolecules, 
spatio-temporal coherence, with superconducting qubits and number
of particles, with {\it e.g.} self interference of single particles in complex molecules. 
An entanglement in  large atomic ensemble (3000~atoms) {\it via} the interaction
with a very weak laser pulse has been observed by a Americano-Serbian team
(McConnell {\it et al.}, 2015).
Quantum spin dynamics and entanglement with hundreds of trapped ions 
is related by the team of  Bohnet (2016).
The quantum brain model of Vitiello (Ricciardi and Umezawa, 1967, 
Del Giudice {\it et al.}, 1988, Vitiello, 1995) 
show an other example of macroscopic entangled state. 
The water inside our brain could be a macroscopic set of correlated matter. 
Much of our understanding of human thinking is based on probabilistic models.
With a quantum calculus of probabilities, Busemeyer and Bruza (2012) 
show that a much better account of human thinking is possible than with traditional models. 
Khrennikov (1999) has proposed classical and quantum mechanics on information spaces 
to understand  anomalous phenomena in cognition,  psychology and sociology.
A quantum-like interference effect in gene expression is studied  in the work 
of Basieva {\it et al.} (2011). 
In their book, Haven and Khrennikov  (2013) explain why
quantum mechanics can be applied outside of physics and define quantum social
science. 
The adaptation of the mathematical formalism of quantum information theory in the biological context
is proposed in the book of Asano {\it et al.} (2015).

\bigskip \bigskip   \noindent {\bf \large    2) \quad  Fractaquantum hypothesis }   

\smallskip
The present fractaquantum idea considering macroscopic applications of quantum mechanics 
is in the same spirit than from previous authors like Heisenberg (1969) 
for pioneering ideas concerning quantum extensions. We refer also to McFadden (2000) 
for biology or Stapp (1993) for mind and brain. Note also intensive development 
concerning violation of Bell inequalities at a macroscopic scale by Aerts et al (2000), 
and Conte et al (2008) among others.  
We first recall fundamental aspects of fractals, develop our point of view 
concerning Atoms and underline the quantum classification between matter and relations.
Then the fractaquantum hypothesis introduces naturally macroscopic bosons.
We discuss also an important point concerning indiscernability
and suggest the notion of weak intrication. 

\monitem   Fractals  

\noindent 
The introduction by Mandelbrot (1975) of the fractal geometry 
has been coupled with a real trouble. The fundamental remark is as follows: 
if we suppose that ``the big is analogous to the little'', 
we obtain of course the straight lines of elementary geometry but also geometrical 
shapes that are absolutely not straight lines, called fractal curves, as the 
popular Peano curve (1890), Von Koch snowflake (1906) or Sierpi'nski triangle (1915). 
A fractal curve has an infinite length and remains unchanged under very simple 
geometric transformations. These self-similar geometrical shapes are present in 
our natural environment with trees, clouds, ferns or cauliflowers among others. 
They are present also in our own body with the detailed structure occupied by the lungs. 
A partial piece of a tree is analogous to the entire tree and this fractal property 
is characteristic of the fact that ``the big is analogous to the little''. 
Let's bare in mind that there is no constraint, 
Nature offers a spatial self-similarity: the ``big'' is analogous to the ``little'', 
even if the corresponding shapes take a complex appearance.

\monitem   Atoms   

\noindent 
Following a vision that comes from the antic Greek culture (see e.g. Salem, 1997), 
in this contribution, an Atom is any natural element whose qualitative properties are modified 
at least in one subset if we divide it into two parts. 
Of course, the modern atoms  of Perrin (1913) 
that are studied with the atomic physics are Atoms in our understanding. 
A fundamental particle, {\it id est}, an entity without any internal relations, 
is an Atom and the  existence of spin as an intrinsic kinetic momentum 
does not imply a relation. In particular, all classical 
elementary particles of microphysics, proton, neutron, and electron, ...   are Atoms. 
Stable structures in Nature such as molecules in the usual sense given in chemistry 
are also Atoms. 
If I divide a glass of water, say 80~times, I have no more water.
The structure of the molecule is broken and I obtain something else. The qualitative properties
of water have been strongly modified by the last division. 
Moreover, the notion of Atom is not reduced 
to the micro-scale and we consider here a living cell as an Atom, due to all 
the properties that are strongly modified or destroyed if it is cut into two parts. 
We extend the family of Atoms to highly organised living organisms, 
including mammals and human beings. At a superior scale, it is not clear for us 
that the entire social organisation of life and exchanges on Earth constitutes or not an Atom, 
as suggested by Lovelock and Margulis (1974).

\monitem   Matter and relations    

\noindent 
The main issue due to quantum theory according to us is the separation between ``matter'' 
and ``relations''. On the one hand, matter is composed with  fermions 
(protons, electrons, etc.) that are indistinguishable and follow the statistics of 
Fermi-Dirac. On the other hand, the relations, {\it i.e.} the interactions between elements of 
matter, are composed by  bosons, like photons, that are the elementary components of light. 
The bosons are also indistinguishable and follow the statistics of Bose-Einstein. 
The indiscernability of identical quantum Atoms is a fundamental postulate 
of the theory that is in clear accordance with the experiments. In particular, 
it is not possible to distinguish between two electrons or between two photons. 
Moreover, the Fermi-Dirac statistics implies also the Pauli exclusion principle 
that claims that two analogous fermions cannot occupy the same position in space. 
Our comment about the Pauli exclusion principle is that ``matter creates space, 
relations  give it a structure''. 

\monitem   Fractaquantum relations  

\noindent 
The fractaquantum hypothesis  (Dubois, 2002) expresses that the quantum framework is pertinent for all 
Atoms in Nature, whatever their size.

\noindent 
Once the fractaquantum hypothesis is taken into consideration, we can explore its consequences 
for simple associations and configurations. Let's bare in mind that the quantum association 
of two identical particles of spin equal to $ \,\sfrac{1}{2} \,$  conducts to a boson of spin equal to zero. 
However, such a boson is a relation because it is a quantum Atom of integer spin. 
Therefore, the anti-symmetric association of two Atoms of matter naturally defines a new relation. 
A fractaquantum consequence of this microscopic property is the existence 
at our scale of a lot of temporal associations that are constructed and exist 
in order to express some relation, some communication, some exchange (Dubois, 2006). 
A  fractaquantum intercessor is 
a structure composed by two Atoms plus a relation between them. 

\smallskip \noindent 
For more complex structures, we can consider set of $n$  vertices (or fermions for our example, 
Atoms of matter) are in relation, interact through a given set of $m$ binary links, 
edges between two vertices. For example, the fractaquantum intercessor of the previous paragraph 
corresponds to $ n = 2$, $m = 1$. More complex picture can be considered and 
the use of graph theory is now classical: in the framework of chemistry (Eigen, 1971), 
biology (Atlan, 1979). For a fractaquantum intercessor  there is no loop 
and this kind of structure can be simply topologically reduced to a simple vertex (Berge, 1969). 

\monitem   Discussion  

\noindent 
In front of these ideas and strong assumptions, several objections and open questions 
can be formulated. A main drawback of the fractaquantum hypothesis is the contradiction 
of quantum indiscernability with macroscopic appearances. Concerning human beings for example, 
it is obvious that ``we are all different''! One could conclude that the fractaquantum 
hypothesis is absurd and one could not consider it anymore. Our motivation 
to go one-step further is firstly motivated by classical philosophical observations 
introduced by Descartes (1641): ``appearances are deceiving''. 
We take some time to develop some doubt. 
 
\smallskip \noindent 
If we refer to the human example, the common points between two human persons 
are much more important than the different ones. The existence of medicine establishes
 empirically this fact, as remarqued by Nunez (2003). 
 Moreover, the explicitation of genomic structure of 
deoxyribonucleic acid in each human cell (Venter {\it et al.}, 2001) show that 
two human deoxyribonucleic acid sequences coincide up to 1 for 10000 parts. 
Even if the single nucloitidic polymorphism is widely studied 
in order to make in evidence local mutations (see {\it e.g.} Zhao {\it et al.}, 2003), 
the first established accounting fact from genomic studies is that two human beings 
have the same sequence of deoxyribonucleic acid up to 99.99 \%! 

\smallskip \noindent 
There exists also circumstances in quotidian life when two persons can be exchanged. 
With the example of a crowd, studied by Le Bon (1895) and Freud (1921), 
one can consider that a new entity is created, where each human being is reduced 
to a very primitive component and develop intense internal relations. 
In his contribution, Freud (1921) describes also social structures as religious 
organisations and army as artificial crowds! In such hirarchical organisation, 
each Atom is a priori exchangeable and is reduced to specific function. 
 
\smallskip \noindent 
As a conclusion of this discussion, 
the scale invariance in Nature is not an exact property.
It is  weakly  broken due to discernability. 
Nevertheless, the fractaquantum hypothesis gives an interesting
point of view to describe basic aspects of Atoms and Relations in Nature. 
It has to be considered as a starting point of a more precise theory. 
Moreover, we have developed  the fractaquantum hypothesis  in several fields 
like   voting (Dubois, 2009, 2014), serendipity (Dubois, 2011), writing (2014), 
cognition (Lambert-Mogiliansky and Dubois, 2015), 
psychology (Lambert-Mogiliansky and Dubois, 2016) and 
meditation (Dubois and Miquel, 2015). 

\monitem   A remark on individuation  suggested by one referee 

\noindent 
Social and biological systems recognise different
individuals but fermions and bosons of the same species are completely interchangeable.
This presents methodological problems. However, quantum modelling could also illustrate
how individuation is achieved in social and biological systems. Is individuation a result
of inherent properties or a result of interactions with the environment? A quantum system
creates roles and the result of this process may come as a surprise to the scientific
community. An example in point is the nuclear shell model. The nucleus was originally
considered as a liquid-drop like system of protons and neutrons. That it actually
arranged itself into shells with very clear roles for protons and neutrons came as a
total surprise. In 1952, a major and very influential handbook on theoretical nuclear
physics commented on the success of the shell theory for the atomic nucleus: ``We are
facing here one of the fundamental problems of nuclear structure which has not yet been
solved'' (Blatt and Weisskopf, 1952, p. 778). In a seminal paper, Gomes {\it et al.} (1957)
solved the riddle and showed how the organisation into shells emerged from properties of
the nuclear force. The surprises from the atomic nucleus illustrates how a quantum
system of interchangeable elements self-organises into a structured system with very
distinct eigenbehaviour.

%

\monitem   Weak intrication 

\noindent 
A natural question is associated with the fractaquantum hypothesis: does entangled matter
exist at a macroscopic scale? 
Is it possible to evidence 
at a macroscopic scale phenomena that show that two apparently distinct objects belong 
in fact to the same Atom? This question is highly difficult. 
%
The notion of intrication is well known at the microscopic scale.
After an interaction of two systems during a given time interval, 
 a new structure is created, even if it is located at several positions. 
For macroscopic Atoms,  there is no explicit evidence that a past 
interaction can induce a true entangled struture in the future. 
The effect of such  entanglement  is  not necessarily dominant, 
as it is the case for the Aspect photons  for example. 
Nevertheless, the consequences of a past interaction could be permanent.
 In this contribution, we  name  ``weak intrication'' such remaining link 
inside a macroscopic  Atom generated by two previous strutures, 
or by one structure that divides itself. 
Two  weakly intricate Atoms occupy two distinct positions and 
have the appearence of a complete independence. 
 Nevertheless, they still have some characteristics of entanglement:
the interaction with one part of the system can produce an immediate
 reaction in the other part, without any explicit communication. 
We develop in the next section the possibility to interpret acupuncture with 
weak intricate structures associated to embryology.

\bigskip \bigskip   \noindent {\bf \large    3) \quad  Acupuncture and embryology   }   

\smallskip
With the help of the fractaquantum hypothesis, we propose to construct links 
between embryology and acupuncture. 
We have suggested at 
the Afscet-And\'e's meeting (Dubois, 2006, 2014) that relations between 
acupuncture points and internal organs could be the sign of the existence of 
macroscopic intricate state. The correlation with the embryologic development 
could be a possibility to evidence the past interaction between the corresponding cells. 
We first recall fundamentals facts about acupuncture, then develop the question of embryology. 
Finally, we  study possible relations between cell division and  intrication. 

\monitem   Acupuncture   

\noindent 
Empirical knowledge developed in China since 3000 years with the acupuncture. 
A 2000 years old classical book is named ``Nei Jing Su Wen'' or 
 ``Huangdi Neijing'' (Inner Classic of the Yellow Emperor), 
 attributed to the mythical emperor Huang Di. 
Remember that acupuncture sets up some relations between the internal organs inside 
the body and some precise locations on the skin that are acupuncture points. 
Of course, these correlations resist to simple explanations through classical 
scientific approaches, even if recent contributions of Pariente {\it et al.} (2005) 
and Nedergaard {\it et al.}   (Goldmann, 2010) create interesting links between 
modern scientific protocols and traditional acupuncture. 
Our hypothesis (2006, 2014), is the following: the meridians of Acupuncture 
follow families of intricate cells. 
There is no present link between these parts of the body. 
The link has to be founded in the past of the corresponding cells, 
in the embryologic development.

\monitem   Embryology   

\noindent 
A single cell develops in a 
time that can be considered as short or long depending on point of view, 
in order to create a complex 
highly organized living being. The interaction with the environment is crucial 
and the way some global information could be presented at the final state 
of embryonic evolution is an open question that seems to be an acceptable possibility. 
It is very interesting in our quest to review the possibilities of inter-changeability 
of two cells during the embryogenesis. As we all know (see {\it e.g.} the book of Gilbert, 2006), 
a complex organism such as a human being comes from a single cell that divides many times 
and particularizes themselves. At a certain step of embryogenic development, 
all the embryonic stem cells (Evans and Kaufman, 1981, Martin, 1981) are identical 
and are a priori interchangeable. After a certain time, they are different, 
they look different and most important, they have been specialized in specifc 
functions in order to promote the development of the entire Atom to the superior scale. 
The understanding of the exact dynamics during the embryogenesis still is an open question. 
Moreover, we consider here that ethical reasons naturally limit the field of scientific research. 
We can also consider this fact as a macroscopic version of the Heisenberg inequalities. 
Following Heisenberg himself (1969), ``we cannot make any observation without 
disturbing the phenomenon under observation''. 

\smallskip
\centerline { \includegraphics [width=.65 \textwidth, angle=0] {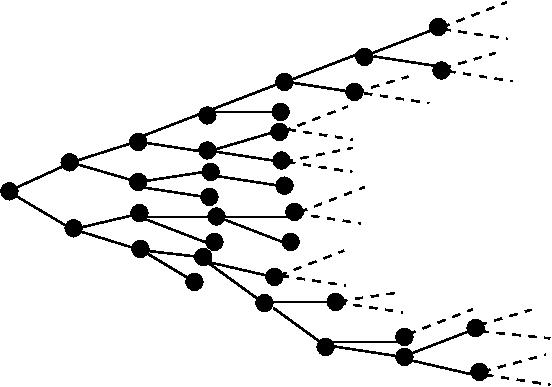} }
\smallskip

\noindent 
Figure 1. Time graph of the daughter cells from a single independent zygote. 
Time is going from left to right. A filiation structure is organized from a 
single zygote (left) to a complex tree. 
\smallskip

\monitem   Cell division and intrication 

\noindent 
The first question concerns  cell division. Could this biological process create 
a macroscopic entangled state ? 
This problem  is not considered with this formulation by 
biological research teams. According to Hatfield {\it et al.}  (2005), 
stem cell division is regulated by the micro ribosomal ribonucleic acid  pathway.
According to Bedzhov and Zernicka-Goetz (2014), 
self-organizing properties of mouse pluripotent cells 
initiate morphogenesis upon implantation.
Nevertheless, according to O'Reilly and Olaya-Castro (2014) or Kourilsky (2016), 
quantum phenomena have to be considered in biology.

\smallskip \noindent 
We place ourselves in a competitive paradigm between 
on the one hand the final unit (quantum holism) and secondly decoherence due to the 
interaction with the environment (``two cells''). Probably an efficient model
of weak intrication  could 
be  intermediate between these two views. In this case, one can imagine that during 
cell division which is the primitive organism, especially during the first cell divisions 
of the blastocyst, a form of global unity, type entanglement, remains persistent.

\smallskip \noindent 
In Figure 1, we stylized the embryonic process as a binary planar graph. It was of course 
a complex dynamic between the two cells ``daughters'' of the same original cell. 
This representation provides a hierarchical breakdown of the body's cells. Indeed, 
a given cell of the human body is ultimately obtained at the end of cell division 
which has not subsequently followed by a new
division. Thus the cells in a way a generation number. We formulated the hypothesis 
that in some sense, the daughter cells remain connected. They form a structure of a tree, 
in the sens of graph theory. Moreover, this tree is an intricate macroscopic state. 
This structure should be naturally universal since embryo development 
is the same for all human beings.

\smallskip \noindent 
Understanding cell division is a kea point concerning the possibility of intrication. 
In this case, there is not a single Atom composed by two components, as in the 
Einstein-Podolsky-Rosen-Bohm experiment. 
At the contrary, a single cell |+> interacts with its environment and generates a double cell 
that we can note as  |++>.
This biological process is absolutly non trivial and we refer to the works of 
Croce and Calin (2005) and Bedzhov and Zernicka-Goetz (2014). From a mathematical 
point of view, quantum field theory has to be introduced since the total mass of the 
refering element (one single cell that becomes two cells) is changed. Probably, 
the co-product of Hopf algebras (see e.g. Cartier, 2006) is a good mathematical tool  
to describe the process of cell-division $ \, |+> \longrightarrow |++> $.

\bigskip \bigskip   \noindent {\bf \large    Conclusion  }   

\noindent 
With the framework of the fractaquantum hypothesis, we have considered 
the possibility of interpreting the acupuncture meridians as intricate cells since 
the early embryologic development. 
What is essential for our purpose, the links between 
the related components of an intricate macroscopic state are not explicit 
through space structure at a given time. It has to be founded in the time process 
of creating the cells. This explains why the meridians are not visible. 
A fundamental question is still open: during the division of a stem cell, 
can some weak form of intrication  be maintained?

\bigskip \bigskip   \noindent {\bf \large    Acknowledgments   }   

The authors  thanks the referee  who suggested several points 
 in need of improvement. Some of them have been included in the final edition
of the contribution.

\bigskip \bigskip      \noindent {\bf  \large  References }   

\smallskip  \noindent \hangindent=7mm \hangafter=1 \noindent 
Aerts D., Aerts S., Broekaert J., Gabora L. (2000), ``The Violation of Bell Inequalities 
in the Macroworld'', {\it Foundations of Physics}, Vol.~53, No~9, pp.~1387--1414. 

\smallskip  \noindent \hangindent=7mm \hangafter=1 \noindent 
Atlan H. (1979),  {\it  Entre le cristal et la fum\'ee, essai sur l'organisation du vivant}, Seuil, Paris. 

\smallskip  \noindent \hangindent=7mm \hangafter=1 \noindent 
Asano M., Khrennikov A., Ohya M., Tanaka Y., Yamato I. (2015), 
 {\it  Quantum Adaptivity in Biology: From Genetics to Cognition}, Springer, New York. 

\smallskip  \noindent \hangindent=7mm \hangafter=1 \noindent 
Aspect A. (2005), Public conference at Orsay University, 02 February 2005. 

\smallskip  \noindent \hangindent=7mm \hangafter=1 \noindent 
Aspect A., Grangier P., Roger R. (1982), ``Experimental realization of Einstein-Podolsky-Rosen-Bohm 
gedanken experiment; a new violation of Bell's inequalities'', {\it Physical Review Letters}, Vol.~49, 
No~2, pp.~91--94. 

\smallskip  \noindent \hangindent=7mm \hangafter=1 \noindent 
Bao X.H., Xu X.F., Li C.M., Yuan Z.S., Lu C.Y., Pan J.W. (2012), 
``Quantum teleportation between remote atomic-ensemble quantum memories'', 
 {\it Proceedings of the National Academy of Sciences of the United States of America}, 
Vol.~109, No~50, pp.~20347--20351. 

\smallskip  \noindent \hangindent=7mm \hangafter=1 \noindent 
Bell J.  (1964).
``On the Einstein Podolsky Rosen Paradox'', 
{\it Physics}, Vol.~1, No~3, pp.~195-200. 

\smallskip  \noindent \hangindent=7mm \hangafter=1 \noindent 
Basieva I., Khrennikov A., Ohya M.,  Yamato I. (2011), 
``Quantum-like interference effect in gene expression: glucose-lactose destructive
interference'', 
{\it Systems and Synthetic Biology}  Vol.~5, pp.~59--68.

\smallskip  \noindent \hangindent=7mm \hangafter=1 \noindent 
Bedzhov I., Zernicka-Goetz M. (2014), ``Self-Organizing Properties of Mouse Pluripotent Cells 
Initiate Morphogenesis upon Implantation'', {\it Cell}, Vol.~156, pp.~1032--1044. 

\smallskip  \noindent \hangindent=7mm \hangafter=1 \noindent 
Bennett C.H.,  Brassard G. (1984), 
``Quantum cryptography: Public key distribution and coin tossing'', 
in Proceedings of IEEE International Conference on Computers, 
Systems and Signal Processing, Bangalore, Vol.~175, pp.~8--12.  
 
\smallskip  \noindent \hangindent=7mm \hangafter=1 \noindent 
Berge C. (1969), {\it Graphes et hypergraphes}, Dunod, Paris. 

\smallskip  \noindent \hangindent=7mm \hangafter=1 \noindent 
Blatt, J.M. and Weisskopf, V.F. (1952), {\it Theoretical Nuclear Physics}, Wiley, New York.

\smallskip  \noindent \hangindent=7mm \hangafter=1 \noindent 
Bohm D. (1951),  {\it Quantum Theory}, Prentice Hall, Englewood Cliffs, New Jersey. 

\smallskip  \noindent \hangindent=7mm \hangafter=1 \noindent 
Bohnet J.G., Sawyer B.C., Britton J.W., 
Wall M.L.,  Rey A.M., Foss-Feig M., Bollinger J.J. (2016), 
``Quantum spin dynamics and entanglement generation with hundreds of trapped ions'', 
 {\it Science}, Vol.~352, No~6291, pp.~1297--1301. 

\smallskip  \noindent \hangindent=7mm \hangafter=1 \noindent 
Branciard C., Gisin N., Kraus B., Scarani V. (2005), 
``Security of two quantum cryptography protocols using the same four qubit states'', 
 {\it Physical Review A},  Vol.~72, No~3, p.~032301.

\smallskip  \noindent \hangindent=7mm \hangafter=1 \noindent 
Brune M., Hagley E., Dreyer J., Maitre X., Maali A., Wunderlich C., Raimond J.M., Haroche~S. (1996), 
``Observing the progressive decoherence of the meter in a quantum measurement'', 
{\it Physical Review Letters}, Vol.~77, pp.~4887--4890.

\smallskip  \noindent \hangindent=7mm \hangafter=1 \noindent 
Busemeyer J.D. , Bruza P.D. (2012), 
{\it Quantum models of cognition and decision}, 
Cambridge University Press,  Cambridge, United Kingdom.

\smallskip  \noindent \hangindent=7mm \hangafter=1 \noindent 
Cartier P. (2006),  ``A primer of Hopf algebras'', Internal report, 
Institut des Hautes Etudes Scientifiques, Bures sur Yvette.

\smallskip  \noindent \hangindent=7mm \hangafter=1 \noindent 
Conte E., Khrennikov A., Todarello O., A. Federici A. (2008), 
``A preliminary experimental verification on the possibility of Bell inequality violation in mental
states''  {\it Neuroquantology},  Vol.~6, No~3, pp.~214--221. 

\smallskip  \noindent \hangindent=7mm \hangafter=1 \noindent 
Croce C.M., Calin C.A. (2005), ``MiRNAs, Cancer, and Stem Cell Division'', {\it Cell},  Vol.~122, No~1, 
pp.~6--7.
 
\smallskip  \noindent \hangindent=7mm \hangafter=1 \noindent 
Debnath S., Linke N.M., Figgatt C., Landsman K.A., Wright K., Monroe C. (2016), 
``Demonstration of a small programmable quantum computer with atomic qubits'', 
 {\it Nature},\br   Vol.~536, pp.~63--66. 

\smallskip  \noindent \hangindent=7mm \hangafter=1 \noindent 
Del Giudice E., Preparata G., Vitiello G. (1988), ``Water as a free electric dipole laser'', 
{\it Physical Review Letters},  Vol.~61, pp.~1085--1088.

\smallskip  \noindent \hangindent=7mm \hangafter=1 \noindent 
Descartes R. (1641), ``Meditationes de prim\^a philosophi\^a, 
ubi de Dei existenti\^a et anim\ae immortalitate'', Mich. Jolly, Paris. 
``M\'editations m\'etaphysiques'', French transation by the duc de Luynes (1647). 

\smallskip  \noindent \hangindent=7mm \hangafter=1 \noindent 
Dubois F. (2002), ``Hypoth\`ese fractaquantique'', {\it Res-Systemica},  Vol.~2, article No.~21, 
available at: www.res-systemica.org/afscet/resSystemica/Crete02/Dubois1.pdf.

\smallskip  \noindent \hangindent=7mm \hangafter=1 \noindent 
Dubois F. (2006), 
``D\'eveloppement, acupuncture et  \'etats macroscopiques intriqu\'es'', 
contribution  presented at the ``Journ\'ees annuelles de l'Afscet'', 
13--14 may 2006, And\'e, available at:
www.afscet.asso.fr/halfsetkafe/textes-2006/acupuncture-2006.html. 

\smallskip  \noindent \hangindent=7mm \hangafter=1 \noindent 
Dubois F. (2006), ``On Fractaquantum Hypothesis'', {\it Res-Systemica},  Vol.~5, article No.~55, 
available at: www.res-systemica.org/afscet/resSystemica/Paris05/dubois.pdf.

\smallskip  \noindent \hangindent=7mm \hangafter=1 \noindent 
Dubois F. (2009), ``On Voting process and Quantum Mechanics'', 
{\it Springer Lecture Notes in Artificial Intelligence}, Vol.~5494 (P.~Bruza {\it et al.} Editors), 
pp.~200--210. 

\smallskip  \noindent \hangindent=7mm \hangafter=1 \noindent 
Dubois F. (2011), 
``Double d\'ecouverte et s\'erendipit\'e'', {\it in}
 {\it La S\'erendipit\'e. Le hasard heureux}, 
Eds D.~Bourcier et P.~van Andel,  pp.~239--247, Hermann, Paris.

\smallskip  \noindent \hangindent=7mm \hangafter=1 \noindent 
Dubois F. (2014), ``On quantum models for opinion and voting intention polls'', 
Proceedings of the 7th International Symposium QI2013, Leicester, 25-27 July 2013, 
{\it Springer Lecture Notes in Computer Science}, 
H.~Atmanspacher, E.~Haven, K.~Kitto, D.~Raine Editors, 
Vol.~8369, pp.~286--295, Springer, New York. 

\smallskip  \noindent \hangindent=7mm \hangafter=1 \noindent 
Dubois F. (2014), ``Acupuncture, embryologie et \'etats macroscopiques intriqu\'es'', 
{\it Res-Syste\-mica},  Vol.~12,  article No.~11, available at: 
www.res-systemica.org/afscet/resSystemica/\br 
vol12-msc/res-systemica-vol-12-art-11.pdf.


\smallskip  \noindent \hangindent=7mm \hangafter=1 \noindent 
Dubois F. (2014), 
``De la dualit\'e sujet-objet \`a la relation observeur-observ\'e'', 
{\it Res-Systemica},  Vol.~12,  article No~12, available at: 
www.res-systemica.org/afscet/resSystemica/vol12- msc/res-systemica-vol-12-art-12.pdf.

\smallskip  \noindent \hangindent=7mm \hangafter=1 \noindent
Dubois F., Miquel C. (2015), 
``Towards a Quantum Model for Meditation'', 
{\it Advances in Systems Science and Applications}, Vol.~15, No~2, pp.~99--119.

\smallskip  \noindent \hangindent=7mm \hangafter=1 \noindent 
Eigen M. (1971), ``Molecular self-organization and the early stages of evolution'', 
{\it Quartely Reviews of Biophysics},  Vol.~4,  No~2--3, pp.~149--212. 

\smallskip  \noindent \hangindent=7mm \hangafter=1 \noindent 
Einstein E., Podolsky B., Rosen N. (1935), 
``Can Quantum-Mechanical Description of Physical Reality Be Considered Complete?'', 
{\it Physical  Review}, Vol.~47, pp.~777--780. 

\smallskip  \noindent \hangindent=7mm \hangafter=1 \noindent 
Ekert A.K. (1991), ``Quantum cryptography based on Bell's theorem''
{\it Physical  Review Letters}, Vol.~67, pp.~661--663. 

\smallskip  \noindent \hangindent=7mm \hangafter=1 \noindent 
Evans M., Kaufman M. (1981), ``Establishment in culture of pluripotential cells 
from mouse embryos'', {\it  Nature},  Vol.~292, pp.~154--156. 

\smallskip  \noindent \hangindent=7mm \hangafter=1 \noindent 
Farrow T., Vedral V. (2015), 
``Classification of macroscopic quantum effects''
{\it  Optics Communications},  Vol.~337, pp.~22--26. 

\smallskip  \noindent \hangindent=7mm \hangafter=1 \noindent 
Feynman R. (1982), ``Simulating physics with computers'', {\it International Journal 
of Theoretical Physics},  Vol.~21,  No~6--7, pp.~467--488. 

\smallskip  \noindent \hangindent=7mm \hangafter=1 \noindent 
Freud S. (1921), {\it Massenpsychologie und Ich-Analyse}, Internationaler Psychoanalytischer 
Verlag, Wien. 

\smallskip  \noindent \hangindent=7mm \hangafter=1 \noindent 
Gilbert S.F. (Editor) (2006), {\it Developmental Biology}, 
eighth Edition, Sinauer Associates Inc, Sunderland, Massachusetts. 

\smallskip  \noindent \hangindent=7mm \hangafter=1 \noindent 
Gomes, L.C., Walecka, J.D. and Weisskopf, V.F. (1957), ``Properties of nuclear matter'',
{\it Annals of Physics}, Vol.~3,   No~3, pp.~241--274

\smallskip  \noindent \hangindent=7mm \hangafter=1 \noindent 
Goldmann N, Chen M., Fujita T., ... Nedergaard N. (2010), ``Adenosine A1 receptors mediate 
local anti-nociceptive effects of acupuncture'', {\it Nature Neurosciences},  Vol.~13, pp.~883--888. 

\smallskip  \noindent \hangindent=7mm \hangafter=1 \noindent 
Hatfield S.D., Shcherbata H.R., Fischer K.A., Nakahara K., Carthew R.W., Ruohola-Baker H. (2005), 
``Stem cell division is regulated by the microRNA pathway'', 
{\it Nature}, Vol.~435, pp.~974-978. 

\smallskip  \noindent \hangindent=7mm \hangafter=1 \noindent 
Haven  E., Khrennikov  A.,  (2013), 
{\it  Quantum Social Science},  Cambridge University Press,  Cambridge,  United Kingdom.

\smallskip  \noindent \hangindent=7mm \hangafter=1 \noindent 
Huangdi Neijing, 
{\it The Su Wen of the Huangdi Neijing} (Inner Classic of the Yellow Emperor),  
Wang Bing's 762 CE version, available at: www.wdl.org/en/item/3044. 

\smallskip  \noindent \hangindent=7mm \hangafter=1 \noindent 
Heisenberg W. (1969), {\it Das Teil and das Ganze, Grespr\"ache im Umkreisis des atomphysik}, 
Piper Verlag, M\"unchen.

\smallskip  \noindent \hangindent=7mm \hangafter=1 \noindent 
Jouguet P., Kunz-Jacques S., Leverrier A., Diamanti E., Grangier P. (2013), 
``Experimental demonstration of long-distance continuous-variable quantum key distribution'', 
{\it Nature  Photonics}, Vol.~7, pp.~378--381. 

\smallskip  \noindent \hangindent=7mm \hangafter=1 \noindent 
Khrennikov A. (1999), 
``Classical and quantum mechanics on information spaces with applications
to cognitive, psychological, social and anomalous phenomena'', 
 {\it Foundations of Physics}, Vol.~29, No~7, pp.~1065--1098. 

\smallskip  \noindent \hangindent=7mm \hangafter=1 \noindent 
Kourilsky P. (2016), 
{\it  Le Jeu du hasard et de la complexit\'e~; la nouvelle science de l'immuno\-logie}, 
Odile Jacob, Paris. 

\smallskip  \noindent \hangindent=7mm \hangafter=1 \noindent 
Lambert-Mogiliansky A., Dubois F. (2015), 
``Transparency in Public Life: A Quantum Cognition Perspective'', 
Proceedings of the 8th International Symposium on Quantum Interaction, 
Filzbach, Switzerland, published in H. Atmanspacher {\it et al.}  (Eds.), 
{\it Springer Lecture Notes in Computer Science}, Vol.~8951, pp.~210--222.

\smallskip  \noindent \hangindent=7mm \hangafter=1 \noindent 
Lambert-Mogiliansky A., Dubois F. (2016), 
``Our (Represented) World: A Quantum-Like Object''
{\it  Contextuality from Quantum Physics to Psychology}, 
Advanced Series on Mathematical Psychology, Vol.~6,  World Scientific, Singapore.   

\smallskip  \noindent \hangindent=7mm \hangafter=1 \noindent  
Le Bon G. (1895), {\it  Psychologie des foules}, Edition F\'elix Alcan, Paris. 

\smallskip  \noindent \hangindent=7mm \hangafter=1 \noindent  
Lovelock J.E., Margulis L. (1974), ``Atmospheric homeostasis by and for the biosphere-- 
The Gaia hypothesis'', {\it Tellus},  Vol.~26, pp.~2--10. 

\smallskip  \noindent \hangindent=7mm \hangafter=1 \noindent  
McFadden J.J. (2000), {\it Quantum Evolution; Life in the Multiverse}, 
Flamingo, Harper \& Collins, London. 

\smallskip \noindent \hangindent=7mm \hangafter=1 \noindent 
Mandelbrot B. (1975),  {\it Les Objets fractals, forme, hasard et dimension}, Flammarion, Paris. 

\smallskip \noindent \hangindent=7mm \hangafter=1 \noindent 
Martin G. (1981), ``Isolation of a pluripotent cell line from early mouse embryos 
cultured in medium conditioned by teratocarcinoma stem cells'', 
{\it Proceedings of National Academy of Science USA},  Vol.~78, pp.~7634--7638.

\smallskip \noindent \hangindent=7mm \hangafter=1 \noindent 
McConnell R., Zhang H., Hu J.,  \'Cuk S., Vuleti\'c V. (2015), 
``Entanglement with Negative Wigner Function of Three Thousand Atoms Heralded by One Photon''
{\it Nature}, Vol.~519, pp.~439--442.

\smallskip \noindent \hangindent=7mm \hangafter=1 \noindent 
Muller A., Zbinden H., N. Gisin N. (1996), 
``Quantum cryptography over 23 km in installed under-lake telecom fibre'',
{\it  Europhysics Letters},   Vol.~33, No~5, pp.~335--339.

\smallskip \noindent \hangindent=7mm \hangafter=1 \noindent 
Nielsen M.A., Chuang I.L. (2000),  {\it Quantum Computation and Quantum Information}, 
Cambridge University Press, Cambridge, UK. 

\smallskip \noindent \hangindent=7mm \hangafter=1 \noindent 
Nunez E.A. (2003), personal communication, Afscet meeting, And\'e, France. 

\smallskip \noindent \hangindent=7mm \hangafter=1 \noindent 
Ofek N., Petrenko A., Heeres R., Reinhold P., Leghtas Z., Vlastakis B., Liu Y., Frunzio L., 
Girvin S.M., Jiang L., Mirrahimi M., Devoret M.H., Schoelkopf R.J. (2016),  
``Extending the lifetime of a quantum bit with error correction 
in superconducting circuits'',   {\it Nature},  Vol.~536, pp.~441--445.

\smallskip \noindent \hangindent=7mm \hangafter=1 \noindent 
O'Reilly E.J., Olaya-Castro A. (2014), 
``Non-classicality of the molecular vibrations assisting exciton energy transfer at room
temperature'',  {\it Nature communications},  Vol.~5, article No.~3012.

\smallskip \noindent \hangindent=7mm \hangafter=1 \noindent 
Pariente J., White P., Frackowiak R., George Lewith G., (2005),  
``Expectancy and belief modulate the neuronal substrates 
of pain treated by acupuncture'',   {\it NeuroImage},  Vol.~25, No~4, pp.~1161--1167. 

\smallskip \noindent \hangindent=7mm \hangafter=1 \noindent 
Peano G. (1890),  ``Sur une courbe qui remplit toute une aire plane,   {\it Mathematische 
Annalen},   Vol.~36, pp.~157--160. 

\smallskip \noindent \hangindent=7mm \hangafter=1 \noindent 
Perrin J. (1913),   {\it Les atomes}, librairie F\'elix Alcan, Paris. 


\smallskip \noindent \hangindent=7mm \hangafter=1 \noindent 
Ricciardi L.M., Umezawa H. (1967),  ``Brain and physics of many-body problems'',  {\it Kybernetik}, 
 Vol.~4, pp.~44--48. 

\smallskip \noindent \hangindent=7mm \hangafter=1 \noindent 
Salem J. (1997),   {\it L'Atomisme antique. D\'emocrite, Epicure, Lucr\`ece}, 
Librairie g\'en\'erale fran\c caise, Paris.

\smallskip \noindent \hangindent=7mm \hangafter=1 \noindent 
Shor P.W. (1994), 
 ``Polynomial-Time Algorithms for Prime Factorization and Discrete Logarithms on a Quantum Computer'', 
Proceedings of the 35th Annual Symposium on Foundations of Computer Science, 
Santa Fe, NM, Nov. 20--22, 1994, IEEE Computer Society Press, pp.~124--134. 

\smallskip \noindent \hangindent=7mm \hangafter=1 \noindent 
Sierpi'nski W. (1915), ``Sur une courbe cantorienne dont tout point est un point de ramification'', 
 {\it Comptes Rendus de l'Acad\'emie des Sciences de Paris},  Vol.~160, pp.~302--305. 
 
\smallskip \noindent \hangindent=7mm \hangafter=1 \noindent 
Stapp H. (1993),   {\it Mind, Matter and Quantum Mechanics}, Springer Verlag, Berlin. 
 
\smallskip \noindent \hangindent=7mm \hangafter=1 \noindent 
Vazirani U., Vidick T. 
(2014), ``Fully Device-Independent Quantum Key Distribution'', 
{\it Physical  Review Letters}, Vol.~113, p.~140501. 

\smallskip \noindent \hangindent=7mm \hangafter=1 \noindent 
Vandersypen L., Steffen M.,  Breyta G., Yannoni C., Sherwood M.H.,  
Chuang I.L. (2001),
``Experimental realization of Shor's quantum factoring algorithm using nuclear magnetic resonance'', 
  {\it Nature},  Vol.~414, No~6866, pp.~883--887. 

\smallskip \noindent \hangindent=7mm \hangafter=1 \noindent  
Venter J.C. {\it et al.} (2001), ``The Sequence of the Human Genome'',  
{\it Science},  Vol.~291, No~5507, pp.~1304--1351.

\smallskip \noindent \hangindent=7mm \hangafter=1 \noindent  
Vitiello G. (1995), ``Dissipation and memory capacity in the quantum brain model'', 
{\it International Journal of Modern Physics-B},  Vol.~9, No~8, pp.~973--989. 

\smallskip \noindent \hangindent=7mm \hangafter=1 \noindent 
Von Koch H. (1904), ``Sur une courbe continue sans tangente, 
obtenue par une construction g\'eom\'etrique \'el\'ementaire'',  
{\it Arkiv f\"ur Mathematik},  Vol.~1, pp~681--704.

\smallskip \noindent \hangindent=7mm \hangafter=1 \noindent  
Xu N., Zhu J., Lu D., Zhou X., Peng X., Du J. (2012), 
 ``Quantum Factorization of 143 on a Dipolar-Coupling NMR system'', 
{\it  Physical Review Letters}, Vol.~108, p.~130501. 

\smallskip \noindent \hangindent=7mm \hangafter=1 \noindent  
Zhao Z. Fu Y.X., Hewett-Hemmett D., Boerwinkle E. (2003), ``Investigating single 
nucloite polymorphism (SNP) density in the human genome 
and its implications for molecular evolution'', {\it Gene},  Vol.~312, pp.~203--213. 

\smallskip \noindent \hangindent=7mm \hangafter=1 \noindent  
Zurek H. (1982), ``Environment induced superselection rules'', {\it Physical Review D}, 
 Vol.~26, No~8, pp.~1862--1880.

\end{document}